\documentclass[final]{aipproc}
\layoutstyle{8x11single}

\usepackage{amsmath,amssymb,amsfonts}
\usepackage{graphicx}
\usepackage[percent]{overpic}

\newcommand{\bb}[1]{{\boldsymbol{#1}}}
\newcommand{\diag}{\mathrm{diag}}
\renewcommand{\d}{\mathrm{d}}
\newcommand{\D}{\mathrm{D}}
\newcommand{\T}{\dagger}
\newcommand{\G}{\mathcal{G}}

\begin{document}

\title{The \textsc{NIFTy} way of Bayesian signal inference}


\author{Marco Selig}
{ address = {Max Planck Institut f\"ur Astrophysik (Karl-Schwarzschild-Stra{\ss}e~1, D-85748~Garching, Germany), and
            Ludwig-Maximilians-Universit\"at M\"unchen (Geschwister-Scholl-Platz~1, D-80539~M\"unchen, Germany)}
}

\begin{abstract}
    We introduce \textsc{NIFTy}, ``Numerical Information Field Theory'', a software package for the development of Bayesian signal inference algorithms that operate independently from any underlying spatial grid and its resolution.
    A large number of Bayesian and Maximum Entropy methods for 1D signal reconstruction, 2D imaging, as well as 3D tomography, appear formally similar, but one often finds individualized implementations that are neither flexible nor easily transferable. Signal inference in the framework of \textsc{NIFTy} can be done in an abstract way, such that algorithms, prototyped in 1D, can be applied to real world problems in higher-dimensional settings.
    \textsc{NIFTy} as a versatile library is applicable and already has been applied in 1D, 2D, 3D and spherical settings. A recent application is the D$^3$PO algorithm targeting the non-trivial task of denoising, deconvolving, and decomposing photon observations in high energy astronomy.
\end{abstract}

\maketitle

\section{Introduction}

    To learn from data is a fundamental concept in science and engineering. In many applications, we want to infer a physical continuous function -- a ``field'' in physicists' terms -- from the measured data.
    In high energy astronomy, for example, X-ray observations are studied for the underlying photon flux distribution in order to deepen our understanding of the causal astrophysical processes. Likewise in the field of medical imaging, X-ray screenings revealing the physical matter density are prescribed due to their diagnostic power. Both, the photon flux and the matter density, are continuous functions of some position space, solid angle on the sky or (projected) spatial position. We denote such a function we are interested in as signal field.

    Before we address the question of how to infer a signal field from a given data set, let us reflect on the conceptual differences between data and signal. Formally, data can be regarded as a finite set of numbers obtained from a measurement. For convenience, we condense all available data in a single data vector $\bb{d} = (d_1,d_2,d_3,\dots)^\intercal$. The dimension of this vector is finite since the measurement, for practical reasons, is of finite resolution and coverage with regard to position, time, energy, and the like. A signal field, however, might vary continuously with respect to those quantities. We assume the signal field $\bb{s} = s(x)$ to be a continuous function on some space $\mho$, with $x \in \mho$. This space $\mho$ can describe a position, time, or energy domain; or even a combination of such domains. The signal field $\bb{s}$ is thereby defined for every $x \in \mho$, and has infinitely many degrees of freedom. In (computational) practice, we need to discretize each field of course for usage within a computer algorithm.

    The inverse problem of constructing an estimate for the signal field from a finite data set is, in general, ill-posed due to the profound signal ambiguity. That means there are (infinitely) many signal field configurations that can lead to the very same data.
    That is because, on the one hand, we lose information on the signal during the measurement process, where an infinite-dimensional quantity is mapped to a finite-dimensional one. For example, fluctuations in the photon flux can only be captured on spatial scales that the observation is able to resolve. On the other hand, the measurement is prone to noise; e.g., Poissonian shot noise in the case of integer photon counts. Even if a measurement of a (unchanged) signal is repeated under identical conditions, a different noise realization can lead to different data.

    The literature on methodological approaches to solve such inverse problems is extensive. Most successful inference schemes are based on Bayesian reasoning using maximum entropy arguments and exploiting (empirically or theoretically motivated) assumptions on likelihood and prior statistics \cite{B63,C46,J57}. One theory, among many other, is Information Field Theory (IFT) \cite{EFK09,E13} that raises Bayesian signal inference to a statistical field theory. The goal hereby remains the derivation of optimal filters -- in the sense of least square error -- that applied to a data vector yield a posterior mean estimate for the signal field. A classical example would be the well-known Wiener filter \cite{W49}, which in the IFT frame corresponds to a free (interactionless) theory.

    Formally, we can derive and formulate inference algorithms in an abstract and mathematical language. Therefore, there is no need to explicitly specify the dimensionality of the problem, and in particular to specify neither the space $\mho$ nor its counterpart in the computer environment. In theory, we could apply a worked out algorithm disregarding the resolution, the total size, and even the geometry of the inference scenario at hand. This kind of universality is part of the success story of inference algorithms like the Wiener filter, which is applied in the most diverse areas.

    Why should we not transfer this universality of inference algorithms to their numerical implementations? The need of and benefit from multi-purpose algorithms that operate independently of the topology and dimensionality of the underlying spaces seems obvious. X-ray imaging techniques developed for high energy astronomy, for example, might also be applicable to medical imaging problems, and vice versa. This would also grant the advantage of prototyping inference algorithms in a simplified setting and of reusing existing codes. Notice that such an algorithm also yields signal field estimates that are independent of the chosen computational grid and its resolution. The reconstruction of a signal field with low and high resolution would look alike up to features that are only representable in the latter.

    All this functionality has now become possible for users resorting to the \textsc{NIFTy} package. \textsc{NIFTy} abbreviates ``Numerical Information Field Theory'', and is publicly available at \url{https://www.mpa-garching.mpg.de/ift/nifty/}.


\section{\textsc{NIFTy}}

    The \textsc{NIFTy} library is written in \textsc{Python} in order to provide the user with a clear and easily understandable interface. Another advantage of the \textsc{Python} interface is that users can make use of the huge number of available packages. These comprise sophisticated numerical libraries as well as application oriented packages; e.g., for astronomical instruments. For computational heavy tasks, libraries written in \textsc{Cython}, C++ and C are used for efficiency.

    The most important property of \textsc{NIFTy} is that it operates regardless of the underlying spatial grid and its resolution. Say we want to synthesize a statistically homogeneous and isotropic Gaussian random field from a (Fourier or angular) power spectrum, for example. Given a concrete power spectrum $p(k)$, we know which amplitudes of fluctuations to expect on certain (spatial, time or energy) scales $x \propto 1/k$. If we draw a random field from this power spectrum, we expect the same kind of fluctuations no matter how we define the space $\mho$. In other words, we could choose $\mho$ to be a one- or two-dimensional grid with low or high resolution, but each random field realization would exhibit similar fluctuations.

    \textsc{NIFTy} makes this possible through an object-orientated framework that offers, among others, classes for spaces, fields, and operators. Here, a space represents a certain type of computational grid. A field, which needs to be defined on some space, primarily carries an array of field values. Operators, which are applied to fields, perform specified (not necessarily linear) transformations.
    All grid dependent computations are hereby forwarded to the corresponding space, where the correct transformations and normalizations are executed.

    In order to emphasize the implied subtlety, let us consider two scalar products. In the data domain, scalar products are defined as in any linear algebra textbook,
    \begin{align}
        \bb{d} \bullet \bb{a} &= \sum_i \; d_i^\ast \cdot a_i
        , \label{eq:sumi}
    \end{align}
    where $\ast$ denotes complex conjugation. The scalar product between signal fields, however, is described by an integral over the continuous space $\mho$,
    \begin{align}
        \bb{s} \bullet \bb{u} &= \int_\mho \d x \;\; s^\ast(x) \cdot u(x)
        . \label{eq:sumx}
    \end{align}
    In order to approximate such integrals by (finite) sums in the computer environment, a proper discretization scheme for $\mho$ needs to be specified. For this purpose, we define a mapping, $s(x) \mapsto s_q$, that discretizes a signal field; e.g., by an evaluation at certain grid positions, $s_q = s(x_q)$. This yields
    \begin{align}
        \bb{s} \bullet \bb{u} &\approx \sum_q \; w_q^{\phantom{\ast}} \cdot s_q^\ast \cdot u_q^{\phantom{\ast}}
        , \label{eq:sumq}
    \end{align}
    where the summation respects the volume elements $\d x$ in form of an appropriately chosen weights $w_q$; e.g., the physical volume of the considered bin, cell, pixel, or voxel indexed by $q$. Those weights are important for the proper normalization, not only of scalar products, but of all operations that involve integrations over $\mho$. A direct consequence of this is that the continuum limit is always preserved. In this way, the resolution independence of the calculus with fields is ensured.

    \textsc{NIFTy} automatizes those normalizations without concerning the user, who simply specifies a space that internally handles all those details. We can choose from a set of preimplemented spaces, which comprises point sets, $n$-dimensional regular grids, spherical spaces, all their harmonic counterparts, and product spaces constructed as combinations of those. The \textsc{NIFTy} toolkit further offers its users a number of generic and exemplary operators, and several useful numerical tools.

    On this basis, \textsc{NIFTy} allows us the abstract formulation and programming of signal inference algorithms. The space $\mho$, on which we define the signal field, is then a mere input for the algorithm. Such an algorithm can run on any space (i.e., regardless of its topology or dimensionality) and with any discretization (i.e., independent of the chosen resolution).

    \begin{figure*}[t]
        \centering
        \includegraphics[width=\textwidth]{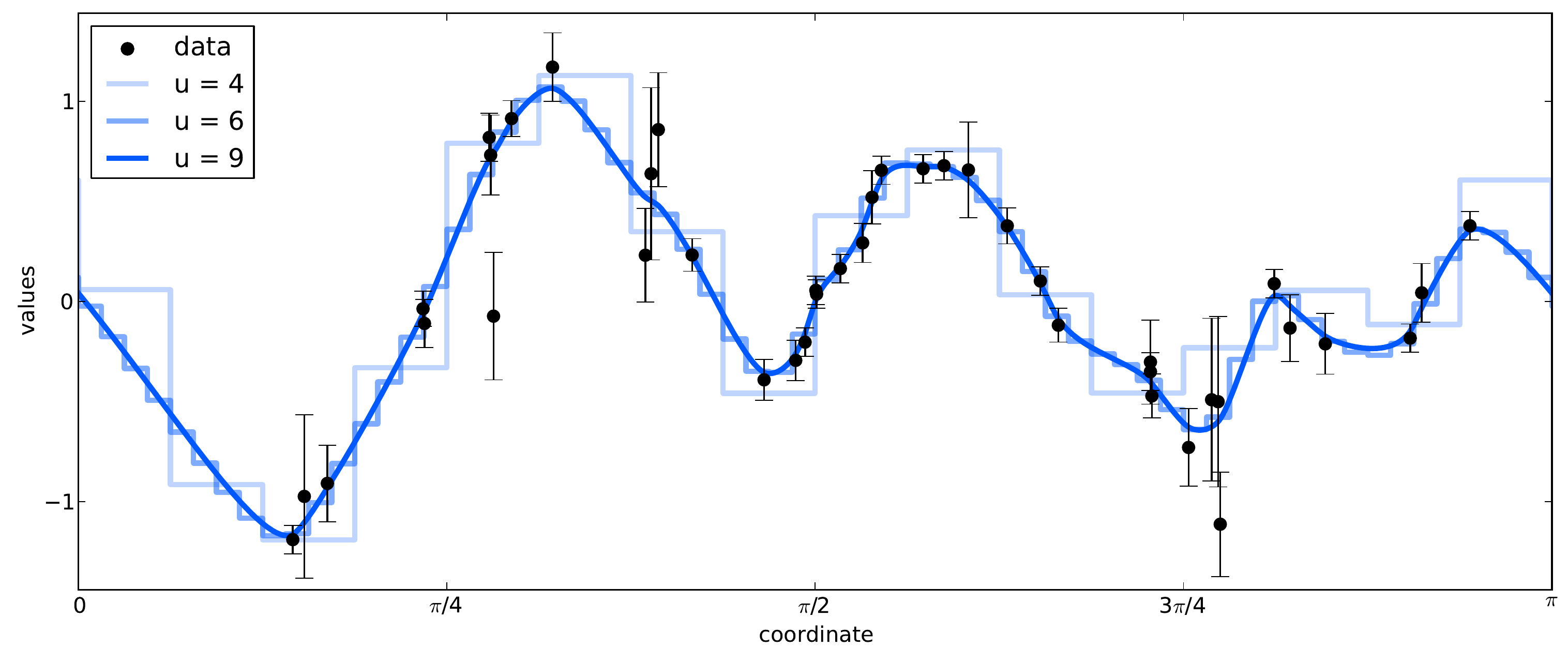}
        \caption{1D signal reconstruction given $42$ noisy data points. The (periodic) position space $\mho = [0,\pi[$ is discretized by $2^\mathrm{u}$ pixels; i.e., $16$, $64$, and $512$ pixels, respectively.}
        \label{fig:42}
    \end{figure*}

    A generic problem is the inference of a (one-dimensional) signal from a set of noisy data points. The signal $\bb{s} = s(x)$ is a function of position $x$ and not constrained by an analytic model.
    Fig.~\ref{fig:42} illustrates an exemplary set of data points $\bb{d}$ and the corresponding (symmetric) error bars $\pm\bb{\sigma}$. The maximum entropy principle thus suggests a zero-mean Gaussian distribution $\G(\bb{n},\bb{N}=\diag[\bb{\sigma}])$ for the noise $\bb{n}$. Given that the data was obtained by a linear measurement process, $\bb{d} = \bb{R}\bb{s} + \bb{n}$, described by a response operator $\bb{R}$, we find the likelihood $P(\bb{d}|\bb{s})$ to be a Gaussian $\G(\bb{d} - \bb{R}\bb{s},\bb{N})$.

    Next, we need to specify our prior assumptions on the signal; i.e., the individual degrees of freedom $s_q$ in the chosen discretization of the position space.\footnote{Neglecting any prior assumptions would result in the so-called matched filter that, in this scenario, would yield a signal reconstruction that is a constant line with spikes at the locations of the data points. Although such a reconstruction would satisfy the likelihood, it contradicts our (\emph{a~priori} and \emph{a~posteriori}) intuition of a good fit.}
    We would \emph{a~priori} expect the signal to be continuous and somewhat smooth; i.e., we anticipate a spatial correlation between the degrees of freedom $s_q$. Again, the maximum entropy principle prompts us to assume a multivariate Gaussian prior $P(s) = \G(\bb{s},\bb{S})$ in this case. For simplicity, we assume the signal covariance $\bb{S}$ to be known. However, in real-life applications, where this is seldomly possible, we would resort to more elaborate methods such as critical filtering \cite{OSBE12,EF11} that allow for the simultaneous reconstruction of $\bb{s}$ and $\bb{S}$ from the same data.

    Computing the posterior distribution according to Bayes' theorem, we obtain another Gaussian distribution. This reduces the signal reconstruction to the task of maximizing the posterior or, equivalently, minimizing its negative logarithm, which is a quadratic form in the signal $\bb{s}$,
    \begin{align}
        -\log P(\bb{s}|\bb{d})
        &= \tfrac{1}{2} (\bb{d}-\bb{R}\bb{s}) \bullet \bb{N}^{-1} (\bb{d}-\bb{R}\bb{s}) + \tfrac{1}{2} \bb{s} \bullet \bb{S}^{-1} \bb{s} + \mathrm{const.}
        \label{eq:wf0} \\
        &= \tfrac{1}{2} \bb{s} \bullet (\bb{R}^\T\bb{N}^{-1}\bb{R} + \bb{S}^{-1}) \; \bb{s} - \bb{s} \bullet (\bb{R}^\T\bb{N}^{-1}\bb{d}) + \mathrm{const.}
        , \label{eq:wf1}
    \end{align}
    where $\T$ denotes adjunction. The resulting \emph{a~posteriori} signal estimate $\left< s \right> = (\bb{R}^\T\bb{N}^{-1}\bb{R} + \bb{S}^{-1})^{-1} (\bb{R}^\T\bb{N}^{-1}\bb{d})$ is well-known as Wiener filter reconstruction. Notice that this solution, as well as the above equations, involve several operations that invoke differently weighted sums according to Eq.~\eqref{eq:sumi} or \eqref{eq:sumq}, respectively.
    With \textsc{NIFTy} doing the proper bookkeeping, we can reformulate this filter as an algorithm and apply it regardless of the chosen discretization.\footnote{The effect of a bookkeeping error corrupts the reconstruction. An incorrect weighting commonly causes a pixelization dependend suppression or overemphasis of the prior with respect to the likelihood, and thus yields reconstructions that dependend on the chosen pixelization.}
    We achieve a resolution independent reconstruction that converges to the continuum limit with increasing resolution. Fig.~\ref{fig:42} shows a few signal reconstructions obtained with the same algorithm for different total numbers of pixels.

    This functionality has been demonstrated with further numerical examples, which successfully applied Wiener filtering on one- and two-dimensional regular grids, as well as spherical \textsc{HEALPix} grids \cite{S+13}.
    However, Wiener filtering is just one example of the versatile applications one can realize using the \textsc{NIFTy} library. A couple of applications are listed in the next section.

    Although \textsc{NIFTy} was originally designed to easily implement algorithms derived in the framework of IFT, it offers a versatile toolkit for any kind of signal inference. Anyone, who wants to infer signal fields from measurement data, can resort to the publically available \textsc{NIFTy} package. \textsc{NIFTy} provides its users an extensive online documentation, tutorials for those new to the subject, and also several demonstration codes. All this (including download and references) can be found on the project homepage at \url{https://www.mpa-garching.mpg.de/ift/nifty/}.

    \begin{figure*}[t]
        \centering
        \begin{tabular}{ccc}
            \begin{overpic} [width=0.32\textwidth]{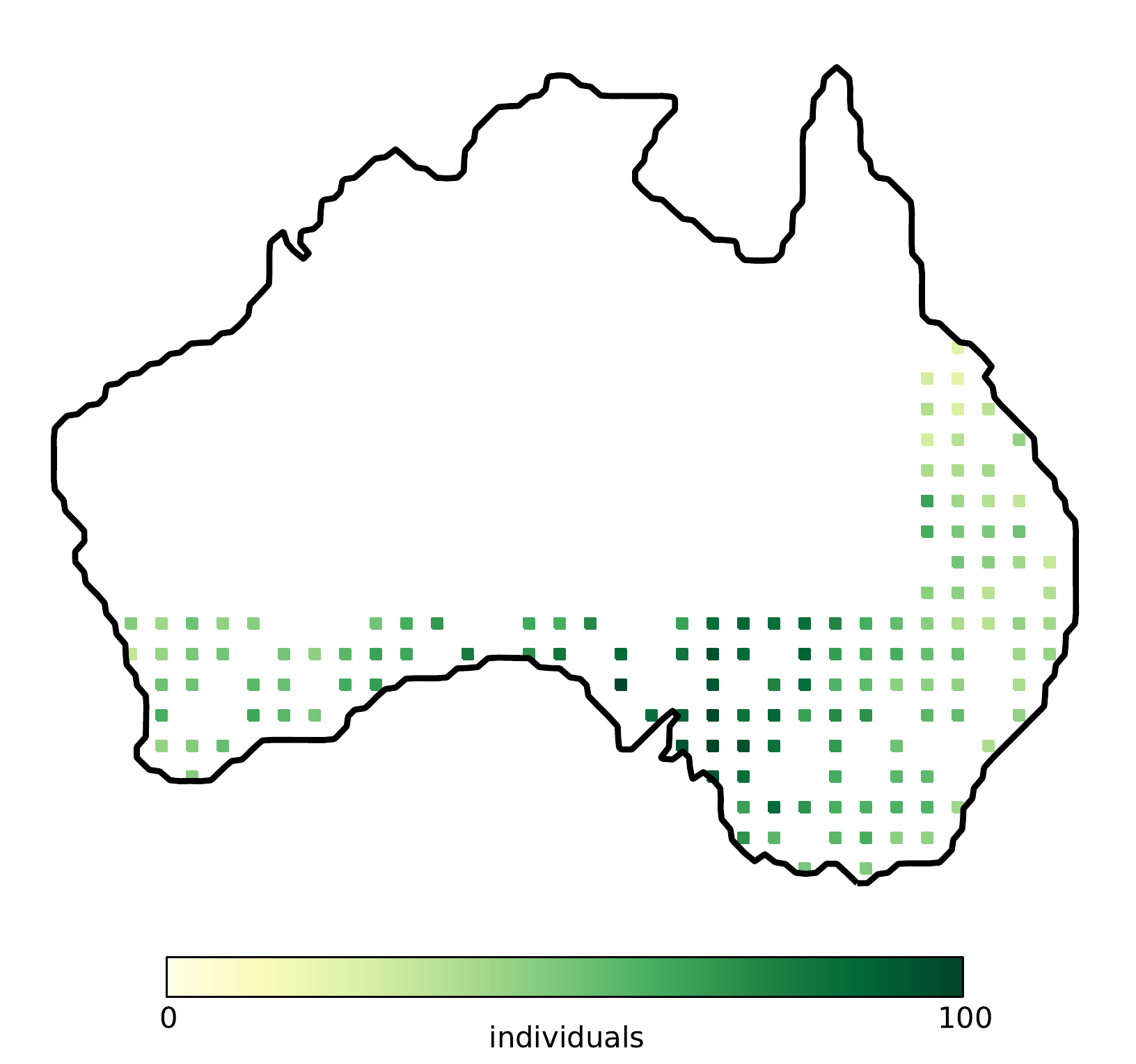} \put(6,82){(a)} \end{overpic} &
            \begin{overpic} [width=0.32\textwidth]{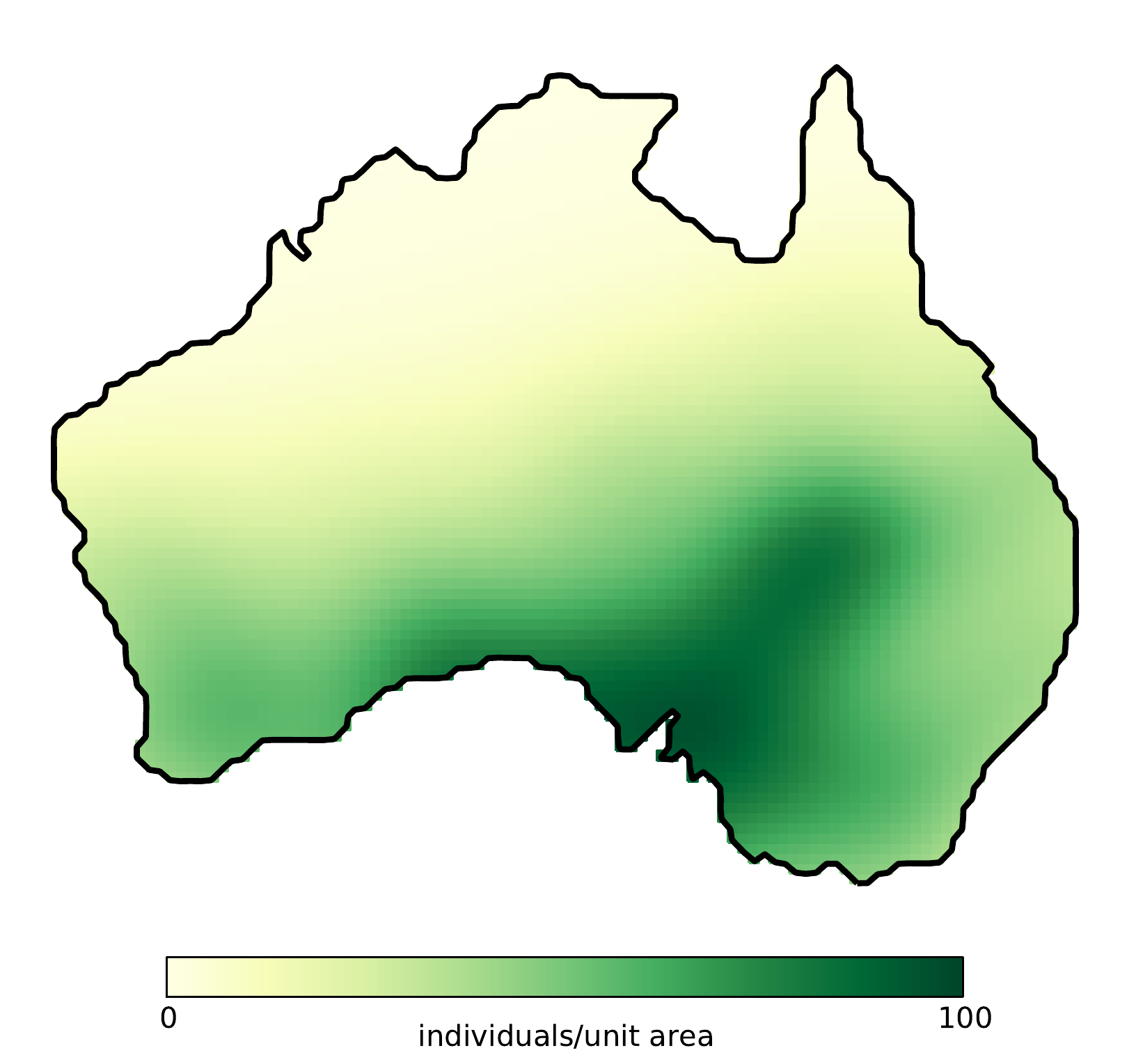} \put(6,82){(b)} \end{overpic} &
            \begin{overpic} [width=0.32\textwidth]{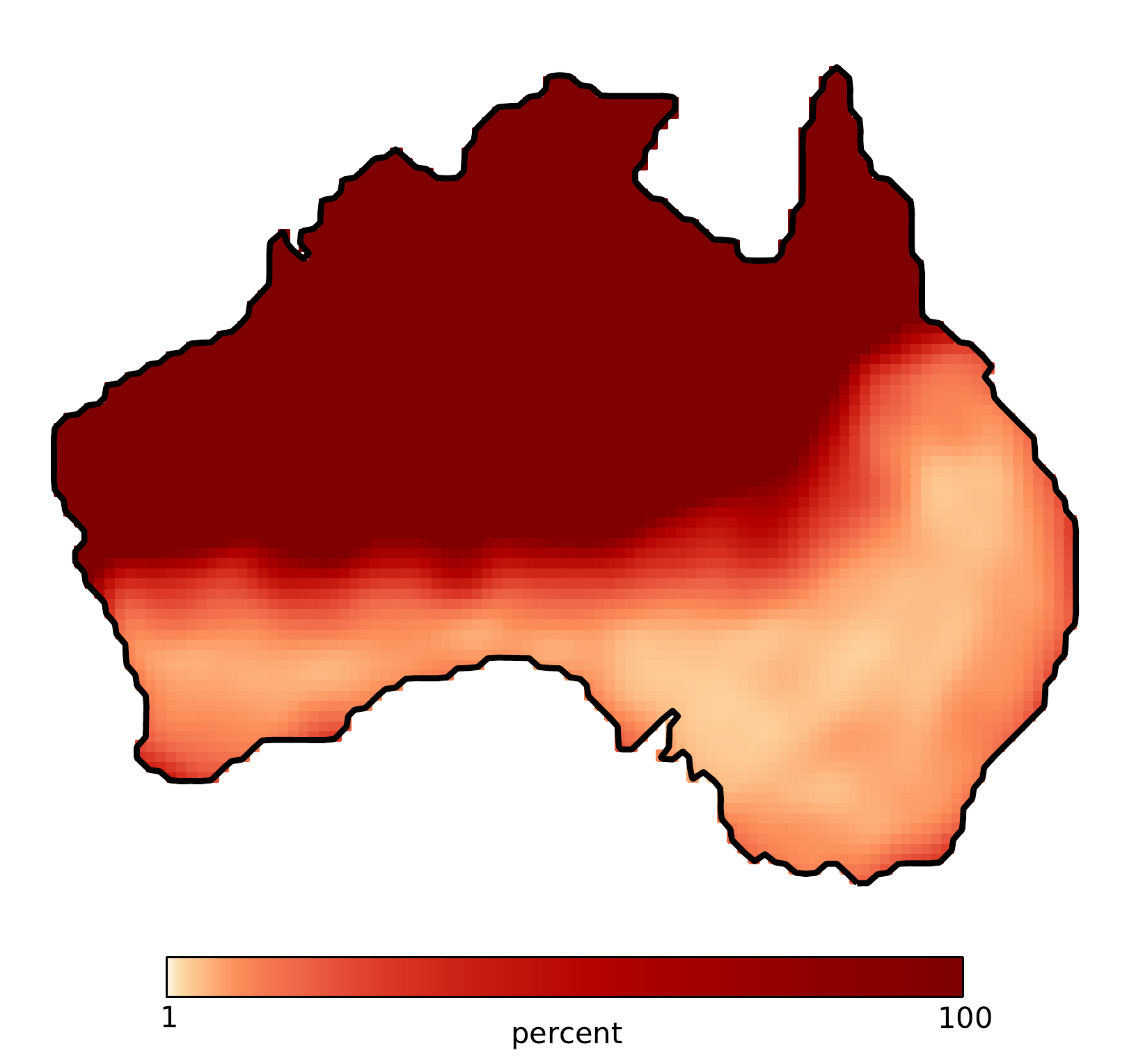} \put(6,82){(c)} \end{overpic} \\
        \end{tabular}
        \caption{2D signal reconstruction with D$^3$PO ``light'': \textbf{(a)} raw data from the census of individuals in a few habitats of unit area, \textbf{(b)} inferred distribution of the species per unit area, \textbf{(c)} relative uncertainty of the inference estimate (cut above $100$\%).}
        \label{fig:au}
    \end{figure*}

\section{Applications}

    \textsc{NIFTy} has already been used in a number of applications, some of which are reviewed here to illustrate its versatility. Subsequent to this listing, we discuss the application of the D$^3$PO ``light'' algorithm to the analysis of census data in detail.

    \textbf{Wiener filtering} is a basic application of Bayesian probability theory, in general, and for IFT in particular, since it represents the free theory around which perturbation series can be expanded. The \textsc{NIFTy} package includes a few Wiener filter code examples serving demonstrative and educational purposes \cite{S+13}.

    \textbf{Log-normal statistics} are often assumed in order to model physical fields that are strictly positive; e.g., the observable photon flux distribution.
    \textsc{NIFTy} has been applied to log-normal models that need to solve non-linear, self-consistent filter equations \cite{OSBE12,SE13,JBSE13,GE13}.

    \textbf{Spectral smoothness priors} for the (Fourier or angular) power spectrum have been studied in the IFT framework \cite{OSBE12,EF11}, and have recently been incorporated into a submodule of the \textsc{NIFTy} package.

    \textbf{D$^\mathbf{3}$PO} is an inference algorithm for Denoising, Deconvolving, and Decomposing Photon Observations in high energy astronomy and related areas of application \cite{SE13}. This complex algorithm tackles the problem of inferring the diffuse and point-like photon flux simultaneously from one raw image of photon counts. The raw image is assumed to be corrupted by imprints of the instrument response functions it has been convolved with, and to suffer from Poissonian shot noise.
    The inference algorithm is derived in the IFT frame and uses a hierarchical Bayesian parameter model for the hyper- and nuisance parameters as well as the signal fields. The denoising (i.e., the suppression of noise) and the deconvolution (i.e., the removal of observational artifacts) are essential steps for signal reconstruction in general. The decomposition task, in which the algorithm allocates flux contributions to the diffuse or point-like component, is the most tricky one. A crucial role plays hereby the exploitation of \textit{a~priori} and \textit{a~posteriori} correlation structures of the signal fields. We assume log-normal prior for the diffuse signal field, a log-uniform and a smoothness prior for its power spectrum, and inverse-gamma priors for the point-like flux.
    Given a data set of photon counts, the D$^3$PO algorithm successfully denoises, deconvolves, and decomposes the raw image into a diffuse and a point-like photon flux reconstruction.
    Currently, photon count data from the INTEGRAL/SPI instrument, the Chandra X-ray observatory, and the Fermi $\gamma$-ray space telescope are being analyzed.

    \textbf{RESOLVE} (Radio Extended SOurces Lognormal deconVolution Estimator) \cite{JBSE13} is a novel algorithm for aperture synthesis imaging of diffuse emission in radio astronomy whose current implementation relies on \textsc{NIFTy}.

    Further applications addressing the non-Gaussianity of the cosmic microwave background \cite{D+13}, self-calibration \cite{EJWS13}, tomography of the Galactic electron density, (weak) gravitational lensing, and computer tomography in the area of medical imaging are in progress.

\subsection{D$^\mathbf{3}$PO ``light''}

    \noindent
    Typical geospatial data in biology is the census of individuals of a certain species within habitats. The counted numbers of individuals within sampled habitats serve as a basis for inferring the species distribution. This is of importance in geo-ecological contexts; e.g., for the (active) conservation of endangered species.
    For practical reasons, census data is limited to a manageable number of rather small habitats that sample the area of interest sparsely. Fig.~\ref{fig:au}a illustrates a raw data set from a simulated census of individuals of some arbitrary species (say the ``opossum entropia maxima'').
    The question of how to extrapolate from such a census to unobserved areas in order to obtain an estimate of the species distribution arises. Since ecological systems are highly complex, we adopt a (strongly) simplified view in the following.

    Our signal field of interest is the species distribution $\bb{\rho}$ as a function of (geographic) position. It is convenient to discretize the considered area in cells that are aligned and coextensive with the surveyed habitats. This way, the mapping between data and signal space becomes intuitive.
    The census data $\bb{d}$, on the other hand, is finite and noisy. The geospatial incompleteness of the census can be described by an exposure that is $0$ or $1$ depending on whether the respective cell has been surveyed or not. Intermediate exposure values might be chosen; e.g., if a habitat was only inspected partly due to inaccessibility, or if there is reason to consider a bias due to nesting or flight behavior of the species. This exposure can be condensed into a diagonal matrix $\bb{E}$ with which we can describe a deterministic relation between the species distribution $\bb{\rho}$ and the expected number of counts of individuals $\bb{\lambda}$,
    \begin{align}
        P(\bb{\lambda}|\bb{\rho}) &= \delta\left( \lambda - \bb{E}\bb{\rho} \right)
        .
    \end{align}
    The collected census data, however, is subject to noise. The noise statistics can be described by a Poisson distribution, since the measurements is based on individual counts of species members; i.e.,
    \begin{align}
        P(\bb{d}|\bb{\lambda}) &= \prod_i \frac{1}{d_i!} \; \exp\left( -\lambda_i + d_i \log \lambda_i \right)
        . \label{eq:poisson}
    \end{align}
    Under those simplistic but adequate assumptions, we can formulate the likelihood of the inverse problem. Its negative logarithm reads,

    \begin{align}
        -\log P(\bb{d}|\bb{\rho}) &= -\log \int\D\lambda \; P(\bb{d}|\bb{\lambda}) \; P(\bb{\lambda}|\bb{\rho})
        = \bb{1} \bullet \bb{E}\bb{\rho} - \bb{d} \bullet \log(\bb{E}\bb{\rho}) + \mathrm{const.}
        , \label{eq:likelihood}
    \end{align}
    where $\bb{1}$ is a constant data vector, and the logarithm is applied componentwise. Notice that this likelihood is local; i.e., there is no information cross-talk between different habitats. This is why a prior model for the species distribution is indispensable for the extrapolation into unobserved areas.

    In order to construct a prior, we have to consider two crucial constraints. Firstly, the species distribution $\bb{\rho}$ is strictly positive. Secondly, there are geospatial correlations between the degrees of freedom that, at lowest order, can be described by a covariance. In this case, the maximum entropy principle suggest a multivariate log-normal prior $P(\bb{\rho})$; i.e.,
    \begin{align}
        -\log P(\bb{\rho}) &= \tfrac{1}{2} (\log \bb{\rho}) \bullet \bb{S}^{-1} (\log \bb{\rho}) + \tfrac{1}{2} \log(\det[\bb{S}])+ \mathrm{const.}
        \label{eq:prior}
    \end{align}
    This prior model is oversimplified lacking a higher order moments and, more importantly, any ecological constraints involving climate, vegetation, food, and predator information, just to mention a few. More realistic priors could be build up using the maximum relative entropy or Bayesian updating \cite{CG06} applied to available data applicable to the considered species.
    However, we want to stress the importance of exploiting spatial correlations and the benefit of \emph{a~posteriori} uncertainty information, rather than presenting a realistic model, and for this purpose such a simplistic prior suffices.
    We choose a convenient parametrization of the \emph{a~priori} unknown covariance $\bb{S}$ that assumes statistical homogeneity and isotropy of the logarithmic species distribution, and further assume Jeffreys and smoothness prior for the introduced hyperparameters according to \cite{OSBE12}.

    Finally, we find a posterior description for the species distribution given census data by combining Eq.~\eqref{eq:likelihood} and \eqref{eq:prior}. Since this posterior is equivalent to the one derived for denoising, deconvolving, and decomposing photon observations in high energy astronomy, we can apply the D$^3$PO algorithm \cite{SE13}, although in a ``light'' version without the complications of point-like contributions.
    Fig.~\ref{fig:au}b shows the reconstructed species distribution. There, it is evident that the (local) information from the census has been propagated to all positions within the area of interested; i.e., an extrapolation inbetween and beyond the inspected habitats. This became possible by exploiting geospatial correlations encoded in the prior covariance.

    At places far away from census habitats, we simply retrieve the prior's zero-mean. However, knowing the noise statistics and reconstructing the prior covariance, the D$^3$PO algorithm can also provide \emph{a~posteriori} uncertainty information. As shown in Fig.~\ref{fig:au}c, the reconstruction error is around a few percent within the region where the census took place, and increases beyond the census' boarders exceeding $100$\% at twice to three times the typical distance between inspected habitats.
    This uncertainty information reveals where we can trust the reconstruction and where additional data is required in order to get more reliable inference results.

\begin{theacknowledgments}

    My thanks go to my co-developers and collaborators, namely Michael R. Bell, Vanessa B\"ohm, Sebastian Dorn, Torsten~A. En{\ss}lin, Mahsa Ghaempanah, Maksim Greiner, Henrik Junklewitz, Niels Oppermann, Carlos Pachajoa, Martin Reinecke, Ga\v{s}per \v{S}enk and Philipp Wullstein for the support of the \textsc{NIFTy} project as well as the insightful discussions and productive comments.
    Furthermore, I thank the ``MaxEnt and Bayesian Association of Australia, Inc.'' for the awarded student support in order to present this work at the ``33rd International Workshop on Bayesian Inference and Maximum Entropy Methods in Science and Engineering'' in 2013.


\end{theacknowledgments}

\newcommand{\prd}{Phys.Rev.D}
\newcommand{\pre}{Phys.Rev.E}

\bibliographystyle{aipproc}

\bibliography{NIFTY}

\end{document}